\documentclass{iau}
\usepackage{graphicx}
\usepackage{natbib}

\title[Rotational Dynamics of Asteroid 4179 Toutatis ] %% give here short title %%
{Revisit of rotational dynamics of Asteroid 4179 Toutatis from Chang'e-2's flyby}

\author[]   %% give here short author list %%
{Yuhui Zhao$^1$ 
, Shoucun Hu$^1$,  Jianghui  Ji$^1$}

\affiliation{
$^1$Key Laboratory of Planetary Sciences, Purple Mountain Observatory, Chinese Academy of Sciences, Nanjing 210008, China; \\
email: {\tt  zhaoyuhui@pmo.ac.cn, jijh@pmo.ac.cn} }

\pubyear{2015}
\volume{318}  %% insert here IAU Symposium No.
\jname{Asteroids: New Observations, New Models}
\editors{Steven Chesley, Alessandro Morbidelli \& Robert Jedicke, eds.}

\begin{document}

\maketitle

\begin{abstract}
This paper presents analysis of the rotational
parameters of Toutatis based on the observational results from Chang'e-2's
close flyby.
The 3-D shape model derived from ground-based radar observation is used to
calculate the 3-1-3 Euler angles at the flyby epoch, which are evaluated to
be $-20.1^\circ\pm1^\circ$, $27.6^\circ\pm1^\circ$ and
$42.2^\circ\pm1^\circ$.
The large amplitude of Toutatis' tumbling attitude is demonstrated to be the result of the large deviation of  the angular momentum axis and the rotational axis.
Two rotational periods are evaluated to be $5.38\pm0.03$ days for rotation about the long axis
and $7.40\pm0.03$ days for precession of the long axis about the angular momentum vector based on Fourier analysis. These results provide a further understanding of rotational state of Toutatis.
\end{abstract}

\firstsection % if your document starts with a section,
\section{Introduction}

Near Earth asteroid 4179 Toutatis is an Apollo-type object and
categorized as a potentially hazardous asteroid (PHA). It has a
highly eccentric orbit, approximately 4:1 resonant with the Earth.
It was firstly discovered in 1934 and the rediscovery took place in
1989. It has closely encountered the Earth every four years during
the last decade, enabling a variety of ground-based assessments.
Radar observations generated by Arecibo and Goldstone reveal the
irregular shape with two lobes and the tumbling,  non-principal axis
(NPA) rotating state of the asteroid \citep{Ostro1995, Hudson1998,
Ostro1999, Ostro2002, Hudson2003}. \cite{Spencer1995} calculated two
rotational periods of the complicated spin state, which were 7.3 and
3.1 days using optical data acquired during 1992 flyby. Meanwhile,
they were determined to be 5.41 days and 7.35 days based on radar
observations \citep{Hudson1995}. \cite{Ostro1995} pointed out that
this tumbling rotation may be primordial. \cite{Ostro1999} used 1996
near-Earth approach radar observations to determine the two major
rotational periods, which were estimated to be $5.376\pm0.001$ days
and $7.420\pm0.005$ days. \cite{Takahashi2013} established a
dynamical model of rotation and calculated the spin state parameters
using radar observations during five flybys from 1992 to 2008.

The first space-based optical observation of Toutatis was obtained
by the second Chinese lunar probe Chang'e-2 at a distance of
$770\pm120~(3\sigma)$ meters at the end of 2012 \citep{Huang2013}.
Over 400 images were captured during the outbound flyby, through
which we estimated Toutatis' orbit, dimensions and orientations.

\section{Methods and Results}

The three-dimensional shape model derived from radar observations
\citep{Busch2012}, as shown in Figure 1a,  is used to  match the
optical image  acquired by Chang'e-2 (Figure 1b) and calculate the
Euler angles at the flyby epoch  \citep{Zhao2014a,Zhao2015}. Taking
the camera's orientation, the Sun-asteroid-spacecraft angle and
render situations into accounts, we rotate the radar shape model
about its principal axes at an interval of $1^{\circ}$ for each of
the three Euler angles and use image matching procedure to obtain
the orientation parameters. The simulated Euler angles are

\begin{equation} %% (1)
\vec{R}=R_z(42.2^\circ)R_x(27.6^\circ)R_z(-20.1^\circ)\vec{r}.
\end{equation}
\begin{figure}[b]
% \vspace*{-2.0 cm}
\begin{center}
 \includegraphics[width=3.4in]{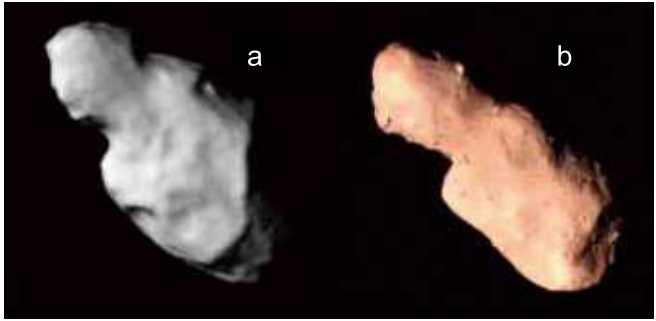}
% \vspace*{-1.0 cm}
 \caption{Images of Toutatis. a: Best-matching attitude of the radar model
with the optical image. b: First panoramic image
captured by Chang'e-2. }
   \label{fig1}
\end{center}
\end{figure}

In combination with radar observations during 1992-2008
\citep{Takahashi2013}, we use the least-squares method to simulate
evolution of rotational dynamics of Toutatis. The external
gravitation torques exerted by the Sun, the Earth and the Moon were
taken into consideration to establish the rotational dynamical
model. The simulated residuals of spin parameters were normalized
with respect to the radar observational errors and they all located
in $3\sigma$ region.

Our simulation results reveal that  the orientation of angular momentum axis could be indicated by
($\lambda_{H}=180.2^{+0.2^\circ}_{-0.3^\circ}$ and
$\beta_{H}=-54.75^{+0.15^\circ}_{-0.10^\circ}$)  in the ecliptic coordinate system, which has remained
nearly unchanged in space over the past two decades. Figure 2 shows the variation of longitude and
latitude of this orientation due to distance from the Sun for the past more than twenty years, the variation amplitudes
are less than one degree in both cases. The slight misalignment of both curves originates from the 2004 near-Earth flyby at a
distance of about 4 Earth-Moon distance. 
The precession amplitude is determined to be
approximately $60^{\circ}$, which results in significantly different attitudes as observed from Earth
at different epochs.

\begin{figure}[h]
% \vspace*{-2.0 cm}
\begin{center}
 \includegraphics[width=3.4in]{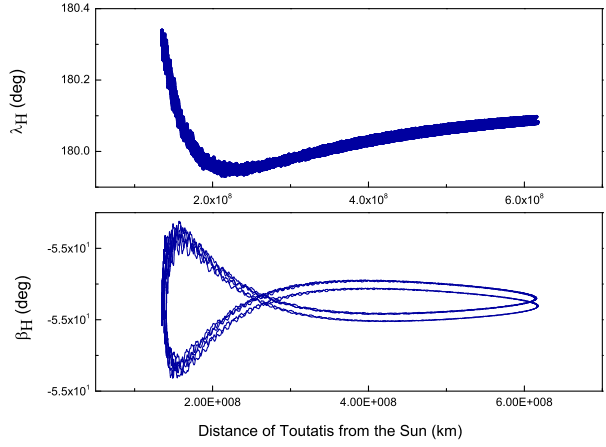}
% \vspace*{-1.0 cm}
 \caption{Variations in Toutatis' angular momentum
orientation from 1992 to 2012. The two panels show the change
in longitude (upper) and latitude (lower) with the distance of Toutatis from the
Sun. }
   \label{fig2}
\end{center}
\end{figure}

Amplitudes of external gravitation torques and the resulting momentum
variations are also simulated in our calculation \citep{Zhao2014b,Zhao2015}
, see also Ji et al 2015, this volume). Solar tides plays
the most important role in the past two decades and near Earth
flybys always led to slight changes.

By applying Fourier transform, the latitudinal variations of the
asteroid's long and middle axes in the J2000 ecliptic frame (as
Figure 3 shows) is simulated to determine the two rotational
periods, which are  5.38 days for the rotation and 7.40 days for
precession \citep{Zhao2015}. These results are in good agreement
with the previous results reported by \cite{Ostro1999}.

\begin{figure}[h]
% \vspace*{-2.0 cm}
\begin{center}
 \includegraphics[width=3.4in]{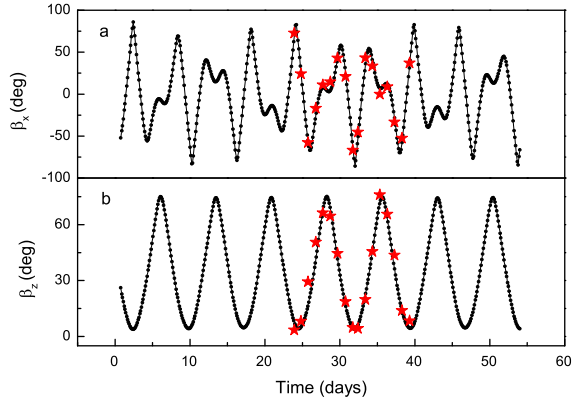}
% \vspace*{-1.0 cm}
 \caption{Latitudinal variations of Toutatis' long axis
$\beta_{z}$ (upper) and middle axis $\beta_{x}$ (lower) in the J2000 inertial
coordinate system in 1992.}
   \label{fig3}
\end{center}
\end{figure}

\section{Conclusions and Discussions}
In this work, we used three-dimensional shape model derived from radar observations to establish the transformation relations between inertial frame, body-fixed frame,  spacecraft frame and instrument's frame. Then we obtained a 3-1-3 Euler angles of the conversion matrix from the celestial coordinate to the body-fixed frame, which are $-20.1^\circ\pm1^\circ$, $27.6^\circ\pm1^\circ$ and
$42.2^\circ\pm1^\circ$. The angular momentum orientation is determined to
be $\lambda_{H}=180.2^{+0.2^\circ}_{-0.3^\circ}$ and
$\beta_{H}=-54.75^{+0.15^\circ}_{-0.10^\circ}$, which is considered to be unchanged for the past two decades.
Fourier analysis is used to obtain the two major rotational periods, which are $5.38\pm0.03$
and $7.40\pm0.03$ days, respectively. Our work provides further investigation of the rotational dynamics of Toutatis.

\section*{Acknowledgements}
This work is financially supported by National Natural Science
Foundation of China (Grants No. 11303103, 11273068, 11473073), the
Strategic Priority Research Program-The Emergence of Cosmological
Structures of the Chinese Academy of Sciences (Grant No.
XDB09000000), the innovative and interdisciplinary program by CAS
(Grant No. KJZD-EW-Z001), the Natural Science Foundation of Jiangsu
Province (Grant No.\\ BK20141509), and the Foundation of Minor
Planets of Purple Mountain Observatory.

\end{document}